\documentclass[12pt]{article}
\pdfoutput=1
\usepackage[top=3cm, bottom=3cm, left=2cm, right=2cm]{geometry}
\usepackage[usenames,dvipsnames,svgnames]{xcolor}
\definecolor{darkblue}{rgb}{0.0,0.1,0.3} 
\definecolor{darkgreen}{rgb}{0,0.65,0}
\definecolor{dblue4}{rgb}{0.06,0.31,0.55} 
\definecolor{nicered}{rgb}{0.7,0.1,0.1}
\definecolor{nicegreen}{rgb}{0.1,0.5,0.1}

\usepackage[numbers,sort&compress]{natbib}

\usepackage[utf8]{inputenc}
\usepackage{textcomp}
\usepackage{amsmath,amssymb,amsfonts,amsthm}
\usepackage{tabularx}
\usepackage{multirow}
\usepackage{dsfont}
\newcolumntype{Y}{>{\centering\arraybackslash}X}

\usepackage{xcolor}
\usepackage[colorlinks=true,citecolor= nicegreen,linkcolor=nicered]{hyperref}

\usepackage{verbatim} 
\usepackage{graphicx}
\graphicspath{{figures/}}
\usepackage{cancel}

\usepackage[colorinlistoftodos]{todonotes}
\usepackage{fontawesome}
\usepackage{comment}
\includecomment{details}
\specialcomment{details}
{\begingroup}{\endgroup}
\excludecomment{details}
\usepackage{tikz}
\newcommand{\ReportNumbers}[1]{%
\begin{tikzpicture}[overlay, remember picture]
\path (current page.north east) ++(-1,-1) node[below left] {#1};
\end{tikzpicture}
}

\title{Dirac neutrino mass generation from Majorana messenger}

\author{Julian Calle${}^1$\footnote{\href{mailto:julian.callem@udea.edu.co}{julian.callem@udea.edu.co}},
  Diego Restrepo${}^{1,2}$\footnote{\href{mailto:restrepo@udea.edu.co}{restrepo@udea.edu.co}}, and Óscar Zapata${}^{1,3}$\footnote{\href{mailto:oalberto.zapata@udea.edu.co }{oalberto.zapata@udea.edu.co}}\\
[2mm]  
\textit{\small ${}^1$Instituto de Física,
  Universidad de Antioquia,
}\\
\textit{\small   Calle 70 \# 52-21, Apartado Aéreo 1226, Medellín, Colombia.}\\
\textit{\small ${}^2$
International Institute of Physics, Universidade Federal do Rio Grande do Norte,
}\\
\textit{\small Campus Universitario, Lagoa Nova, Natal-RN 59078-970, Brazil. }\\
\textit{\small ${}^3$
Abdus Salam International Centre for Theoretical Physics,
}\\
\textit{\small  Strada Costiera 11, 34151, Trieste, Italy. }
}
\date{\small \today}
\begin{document}
\maketitle
\ReportNumbers{IIPDM-2019}

\begin{abstract}    
  The radiative type-I seesaw has been already implemented to explain the lightness of Majorana neutrinos with both Majorana and Dirac heavy fermions, and the lightness of Dirac    neutrinos with                   Dirac heavy fermions. In this work we present a minimal implementation of the radiative type-I seesaw with light Dirac neutrinos and heavy Majorana fermions. An inert doublet and a complex singlet scalar complete the dark sector which is protected by an Abelian fermiophobic gauge symmetry that also forbids tree level mass contributions for the full set of light neutrinos. A fermion vector-like extension of the model is also proposed where the light right-handed neutrinos can thermalize in the primordial plasma and the extra gauge boson can be directly produced at colliders.
In particular, the  current upper bound on $\Delta N_{\text{eff}}$ reported by PLANCK points to large ratios $M_{Z'}/g'\gtrsim 40\ \text{TeV}$ which can be competitive with collider constraint for $g'$ sufficiently large in the ballpark of the Standard Model values, while future cosmic microwave background experiments may probe all the no minimal models presented here. \href{https://github.com/restrepo/DiracMajorana}{\faGithub}
\end{abstract}

\section{Introduction}
\label{sec:intro}

The interpretation of neutrino experimental data in terms of neutrino
oscillations is compatible with both Majorana and Dirac neutrino masses
\cite{Tanabashi:2018oca}. The former possibility has received the most
attention but, given the lack of signals in neutrinoless double beta
decay experiments
\cite{KamLAND-Zen:2016pfg,Agostini:2018tnm,Aalseth:2017btx,Alduino:2017ehq,Albert:2017owj,Arnold:2016bed},
the latter cannot be dismissed.
If neutrinos are Dirac particles, the Standard Model (SM) particle
content must be extended with right-handed neutrinos,
which can increase the effective number of light neutrinos,
$N_{\text{eff}}$, until 6. Therefore, to be compatible with the
current cosmological restrictions on $N_{\text{eff}}$, the
interactions of the extra right-handed neutrinos with the primordial
plasma must be highly suppressed.

On the other hand, to give small masses to at least two Majorana or
Dirac neutrinos, as required to explain the neutrino oscillation
experiments~\cite{Ahmad:2002jz, Fukuda:1998mi},
the seesaw mechanism with heavy fermions is usually invoked.
For the tree-level type-I seesaw we can have either light Majorana
neutrinos with heavy Majorana
mediators~\cite{Minkowski:1977sc,Yanagida:1979as,GellMann:1980vs,Mohapatra:1979ia}
or light Dirac neutrinos with heavy Dirac
mediators~\cite{Roncadelli:1983ty,Roy:1983be,Gu:2007mc,Ma:2014qra}.
The radiative type-I seesaw includes both~\cite{Ma:2006km}
possibilities~\cite{Farzan:2012sa},
but now it is also possible to have light Majorana neutrinos with
heavy Dirac mediators~\cite{Ma:2013yga}.
In this work we want to explore the possibility to build a simple
Dirac radiative type-I seesaw model with heavy Majorana mediators.
It is worth noticing that this idea have been already illustrated in
an extension of the minimal supersymmetric standard
model~\cite{Demir:2007dt} but without show any explicit solution.

In general, solutions for light Dirac neutrino masses require a
continuous symmetry to guarantee their Diracness. This symmetry is
usually identified as the local $\operatorname{U}(1)_{B-L}$.
Additionally, ad-hoc discrete symmetries are invoked to forbid tree
level Dirac or Majorana mass terms for the light right-handed
neutrinos~\cite{Roncadelli:1983ty,Han:2018zcn,Wang:2017mcy}.
However, tree-level Dirac type-I seesaw with proper choices for the
$\operatorname{U}(1)_{B-L}$ charges have been shown to be consistent
without require any extra ad-hoc discrete symmetries~\cite{Ma:2014qra}.
In recent works, it has been shown that even for one-loop Dirac
neutrino masses, it is possible to have $\operatorname{U}(1)_{B-L}$ as
the only extra symmetry beyond the SM~\cite{Calle:2018ovc,Bonilla:2018ynb,Saad:2019bqf}\footnote{
  For extensions with only extra scalars, minimal solutions have been found with two and three loops~\cite{Saad:2019bqf}.}.

As a bonus in this case, the new scalars and fermions circulating the
loop can be dark matter candidates with the stability of the lightest
of them guaranteed by the very same continuous symmetry.
We focus here in solutions for the radiative Dirac type-I seesaw with
Majorana mediators, which have only an extra local symmetry responsible for the
Diracness of the light neutrinos, the absence of any tree-level
mass, and the existence of a dark sector constituted by the
particles circulating the loop.

In fact, another evidence that the SM is not a complete theory is the
missing matter content of the Universe, which is known as dark matter
(DM).
The main proposals that explain DM as a particle are given in
Ref.~\cite{Bertone:2004pz}.
However, there has been only gravitational evidence for the existence
of dark matter so far.
Without evidence of DM as a particle, there is not a clear path to pin
out the DM properties nor the possible heavier companions of some
extended dark sector. 
Linking the dark sector to other specific phenomenology allows to
reduce the arbitrariness in the model building.
In our construction, the dark sector is related to the heavy sector
responsible of the lightness of the neutrinos and the same symmetry
that guarantees the lightness of the Dirac neutrinos is the
responsible of the stability of the lightest dark particle (LDP).
Therefore, the number of specific models is quite restricted.

The rest of the paper is organized as follows. In the next Section we present the model and study the scalar mass spectrum after spontaneous symmetry breaking. In Sec.~\ref{sec:Neutrinos} we present the radiative mechanism that generates Dirac neutrino masses and establish  the lepton flavor violation constraints. The different resulting DM scenarios  are discussed in Sec.~\ref{sec:DM}, and in Sec.~\ref{sec:CosmoConstraints} we show the cosmological restrictions ($N_{\text{eff}}$) in a non-minimum model for different extra Abelian symmetries.

\section{The model}
\label{sec:Model}
\begin{table}[t!]
  \centering
  \begin{tabular}{|c|c|c|c|}
    \hline  
    Fields     & $\operatorname{SU}(2)_L$ & $\operatorname{U}(1)_Y $ & $\operatorname{U}(1)_{\mathcal{D}}$ \\ \hline
    $\eta$  & $\boldsymbol{2}$ & $1$  & $1$ \\
    $S$ & $\boldsymbol{1}$ & $0$  & $2$ \\
    $\sigma$ & $\boldsymbol{1}$ & $0$ & $3$ \\
    \hline
    $\nu_{Ri}$ & $\boldsymbol{1}$ & $0$ & $-4$\\
    $\nu_{R3}$ & $\boldsymbol{1}$ & $0$ & $5$\\
    $\psi_{R\alpha}$  & $\boldsymbol{1}$ & 0 & $1$ \\\hline
  \end{tabular}
  \caption{The new scalars and fermions with their respective charges. All the SM fields are neutral under the dark $\operatorname{U}(1)_{\cal {D}}$ gauge symmetry. }
    \label{tab:partcont}
\end{table}
We extend the SM with a spontaneously broken Abelian gauge 
symmetry which guarantees the diracness of the massive neutrinos.
Only the new particles, including the  right-handed partners of the SM
neutrinos, are charged under this new
$\operatorname{U}(1)_{\mathcal{D}}$ dark gauge
symmetry~\cite{Campos:2017dgc,Bertuzzo:2018itn,Bertuzzo:2018ftf} to obtain an anomaly free theory. We
choose the new particle set such that the following dimension six
operator is realized at one-loop level
\begin{align}
  \label{eq:ld6}
  \mathcal{O}_{6D}=\frac{1}{\Lambda^2} \overline{L} \tilde{H} \nu_R S^2\,,
\end{align}
where $S$ is the singlet scalar field which spontaneously breaks the $\operatorname{U}(1)_{\cal {D}}$ symmetry needed to forbid the Dirac and Majorana neutrino mass terms at tree level.

With the aim to illustrate the one-loop Dirac neutrino mass generation
we consider the particle content shown in Table \ref{tab:partcont} as
a possible realization of the effective operator $\mathcal{O}_{6D}$. Specifically, we introduce three scalar fields
$\eta, \sigma$ and $S$, where only $S$ develops a nonzero vacuum
expectation value (VEV),  a set of three singlet fermions,
$\nu_{Rj}$ ($j=1,2$) and $\nu_{R3}$, and another set of three heavy Majorana fermions, $\psi_{R\alpha}$ ($\alpha=1,2,3$).  
The $\operatorname{U}(1)_{\mathcal{D}}$ charges for the new particles are defined by the anomaly cancellation conditions and the gauge invariance in Yukawa and scalar interactions.

The most general Lagrangian for some gauge $\operatorname{U}(1)_X$ symmetry, which includes the trivial case $X=\mathcal{D}$, must contains the following gauge, Yukawa and scalar interactions in
order to realize $\mathcal{O}_{6D}$ at one-loop:
\begin{align}
\label{eq:LagY}
    \mathcal{L} \supset& -\,g^{\prime}\,Z_\mu^\prime\sum_{F}q_{F}\overline{F} \gamma^\mu F+\sum_{\phi}\left|\left( \partial_\mu +i\,g^{\prime}\,q_\phi\,Z'_\mu \right) \phi\right|^2\nonumber\\
    &-[ 
    h_{i\alpha} \overline{L_{i}} \tilde{\eta} \psi_{R\alpha} +  y_{j\alpha} \overline{\nu_{R_{j}}} \sigma^* \psi^c_{R\alpha} + \kappa_{\alpha\beta} \overline{\psi^{c}_{R\alpha}} \psi_{R\beta} S^* + \text{h.c.}] - \mathcal{V}(H, S, \eta, \sigma)\,.
\end{align}
In the first row $g^{\prime}$ is the gauge coupling associated to the $\operatorname{U}(1)_X$ group and $Z_\mu^\prime$ is its corresponding gauge boson, $F$ ($\phi$) denote the new fermions (scalars), and $q_{F,\,\phi}$ their $X$ charges. In the second row
$L_{i}$ ($i=1,2,3$) and $H$ are the SM lepton and Higgs doublets, respectively,  $\widetilde{\eta} = i \sigma_2 \eta^*$, and $h$, $y$ and $\kappa$ are matrices in the flavor space. 
The scalar potential can be cast as
\begin{align}
  \label{eq:VHSetasigma}
    \mathcal{V}(H, S, \eta, \sigma) = & V(H) + V(S) + V(\eta) + V(\sigma) \nonumber\\
    &+  \lambda_{1} (H^{\dagger} H ) (S^{*} S) + \lambda_{2} (H^{\dagger} H ) (\sigma^{*} \sigma ) + \lambda_{3} (H^{\dagger} H ) (\eta^{\dagger} \eta )\nonumber\\
    &+ \lambda_{4} (S^{*} S) (\sigma^{*} \sigma ) + \lambda_{5} (S^{*} S) (\eta^{\dagger} \eta ) + \lambda_{6} (\eta^{\dagger} \eta ) (\sigma^{*} \sigma ) + \lambda_{7} (\eta^{\dagger} H ) (H^{\dagger} \eta ) \nonumber\\
    &+ \lambda_{8} (\eta^{\dagger} H S^{*} \sigma + \text{h.c.})\,,
\end{align}
with $V(\omega) = \mu^{2}_{\omega} \omega^{\dagger} \omega + \lambda_{\omega} (\omega^{\dagger} \omega)^{2}$. It is worth to emphasize that $\operatorname{U}(1)_{\mathcal{D}}$ automatically allows for all the terms in Eqs.~\eqref{eq:LagY} and~\eqref{eq:VHSetasigma} but other realizations will be checked later on.
Note that after the spontaneous symmetry breaking of $\operatorname{U}(1)_X$ the $\lambda_8$ term gives rise to the mixing between the neutral parts of $\eta$ and $\sigma$, which is mandatory to generate non-zero radiative masses. 
We assume $\lambda_8$ and $\langle S\rangle$ reals to preserve CP symmetry in the scalar sector, and $\mu^2_\eta,\mu^2_\sigma>0$ to avoid tree-level mixing terms among the fermions. 
Moreover, we also assume $\lambda_1\ll1$  such that the scalar $S$ and $H$ do not mix allowing us to identify the CP even scalar particle in $H$ as the SM Higgs boson. 
To establish the scalar spectrum we expand the scalar fields as
\begin{align*}
  H = \begin{pmatrix}G^+ \\ \frac{1}{\sqrt{2}} (h+v_H+iG) \end{pmatrix} \,,&\hspace{1cm}
  \eta = \begin{pmatrix}\eta^{+} \\ \frac{1}{\sqrt{2}}(\eta_R+i\eta_I) \end{pmatrix} \,,\\
  S = \frac{1}{\sqrt{2}} (S_R+v_{S}+iS_I)\,,&\hspace{1cm}\sigma = \frac{1}{\sqrt{2}} (\sigma_R+i\sigma_I),
\end{align*}
with $v_H= 246.22$ GeV.  
Of the original twelve scalar degrees of freedom in the model, the gauge bosons $W^{\pm}$, $Z^{0}$ and $Z^{\prime}$ absorb four of them (the Goldstone bosons $G^{\pm}$,$G$ and $S_{I}$). Thus, the scalar spectrum contains two sets of two neutral CP-even states ($h$ and $S_{R}$,  and $\sigma_{R}$ and $\eta_{R}$), two CP-odd scalar states ($\sigma_{I}$ and $\eta_{I}$) and one charged scalar ($\eta^{\pm}$). 
The mass spectrum for the unmixed scalars reads
\begin{align*}
   & m_{\eta^{\pm}}^{2}= \mu_{\eta}^{2} + \frac{1}{2} (\lambda_{3} \upsilon^{2}_{H} + \lambda_{5} \upsilon^{2}_{S} )\,,\,\,\,\, m_{H}^{2} = \lambda_{H} \upsilon_{H}^{2}\,,\,\,\,\,    m_{S}^{2}= \lambda_{S} \upsilon_{S}^{2}\,.
\end{align*}
The other mass eigenstates are defined as
\begin{align*}
    \begin{pmatrix}\chi_{(R,I)_{1}} \\ \chi_{(R,I)_{2}} \end{pmatrix} =& \begin{pmatrix} \cos\theta & -\sin\theta \\ \sin\theta & \cos\theta \end{pmatrix} \begin{pmatrix}\sigma_{(R,I)} \\ \eta_{(R,I)} \end{pmatrix} \,,
\end{align*}
where $\tan\theta= 2c/(b-a)$, with $a = m^{2}_{\eta^{\pm}} + \frac{1}{2}\lambda_{7} \upsilon^{2}_{H}$, $b = \mu^{2}_{\sigma}+\frac{1}{2}(\lambda_{2} \upsilon^{2}_{H} + \lambda_{4} \upsilon^{2}_{S})$ and $c = \frac{1}{2}\lambda_8 v_{H} v_{S}$\,. Note that the CP-even  states $\chi_{R(1,2)}$ are mass degenerate with CP-odd ones $\chi_{I(1,2)}$, with masses $ m_{\chi_{(1,2)}}^{2}= [a + b \mp \sqrt{(a-b)^{2} + 4c^{2}}]/2$.

On the other hand, we assume that the heavy Majorana fermions are already in the diagonal basis in such a way their masses are $M_{\psi_\alpha}=\kappa_{\alpha\alpha}v_S/\sqrt{2}$, with $M_{\psi_1}<M_{\psi_2}<M_{\psi_3}$. Finally, the mass of the new gauge boson is given by 
\begin{align}
    M_{Z'}=q_Sg^{\prime} v_S.
\end{align}

\section{Neutrino masses and charged lepton flavor violation}
\label{sec:Neutrinos}
\begin{figure}
\centering
\includegraphics[scale=0.6]{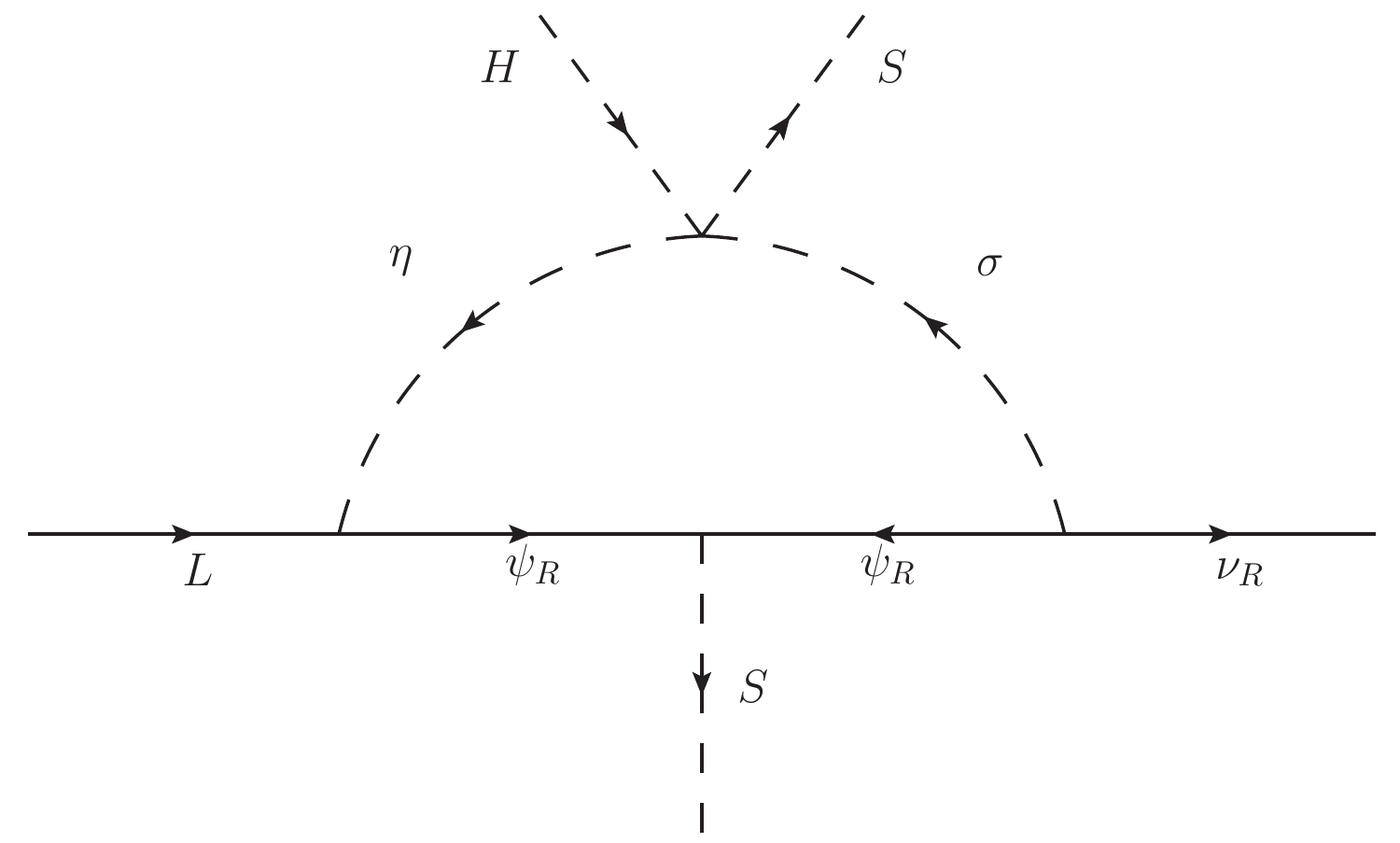}
\caption{One-loop realization of the dimension-6 operator $\overline{L} \tilde{H} \nu_R S^2$ leading to Dirac neutrino masses with Majorana mediators.}
\label{fig:zee}
\end{figure}
Neutrino masses are generated at one-loop level according to the diagram in  Fig.~\ref{fig:zee}. The expression for the effective neutrino mass matrix $\mathcal{M}_{\nu}$ can be cast as
\begin{align}
(\mathcal{M}_{\nu})_{ij} = \frac{1}{32 \pi^{2}}  \frac{\lambda_8 v_S v_H} {m_{\chi_{R_2}}^{2}-m_{\chi_{R_1}}^{2}}\sum_{\alpha=1}^{3} h_{i \alpha} M_{\psi_\alpha} y^{*}_{j\alpha}\left[ F\left( \frac{m_{\chi_{R_2}}^{2}}{M_{\psi_{\alpha}}^{2}} \right) - F\left( \frac{m_{\chi_{R_1}}^{2}}{M_{\psi_{\alpha}}^{2}} \right) \right] + (R \to I)\,,
\end{align}
where $F(x) =x \log x/(x-1)$. 
Note that the structure of the effective neutrino mass matrix, given by the
product $(M_{\nu})_{ij} \propto h_{i \alpha} y_{j\alpha}$, is similar to
the structure of the neutrino mass matrix for the tree-level seesaw
mechanism for Dirac neutrinos~\cite{Chulia:2016ngi}. 
It is also worth mentioning that if only one
fermion $\psi_R$ is added, then there will be two massless neutrinos, which
would be ruled out by the current neutrino oscillation data~\cite{deSalas:2017kay}. 
In our case, we assume the
existence of three of such fermions, generating Dirac scotogenic masses for
the two left-handed neutrinos ($\nu_{3}=\nu_{L3}+\nu_{R3}$ is massless due to the charge assignment). 

In order to estimate the possible values for the parameters involved in the neutrino masses we consider the case where $\lambda_2$, $\lambda_4$, $\lambda_7 \ll 1$ and $m_\chi^2\equiv m_{\eta^{\pm}}^2 = \mu^2_{\sigma} \gg \frac{1}{2} \lambda_8 v v_S$, which leads to $a\approx b\gg c$. 
Taking into account that for the mentioned case $m^2_{\chi_{R_{2}}}-m^2_{\chi_{R_{1}}} = \lambda_8 v v_{S}$ and $m^2_{\chi_{R_{2}}}+m^2_{\chi_{R_{1}}} = 2 m^2_\chi$, we have that
\begin{align}
(\mathcal{M}_{\nu})_{ij} = \frac{\lambda_8 v_S v}{16 \pi^{2}} \sum_{\alpha=1}^{3} \frac{h_{i \alpha} M_{\psi_{\alpha}} y^{*}_{j\alpha}} {m_\chi^{2}-M_{\psi_{\alpha}}^{2}} \left[1 - \frac{M_{\psi_{\alpha}}^{2}}{m_\chi^{2}-M_{\psi_{\alpha}}^{2}} \log \left( \frac{m_{\chi}^{2}}{M_{\psi_{\alpha}}^{2}} \right)\right]\,,
\end{align}
and by further assuming $m_{\chi}^{2} \gg M_{\psi_{\alpha}}^2$ one finds 
\begin{align}
(\mathcal{M}_{\nu})_{ij} & = \frac{\lambda_8 v_S v}{16 \pi^{2}m_{\chi}^{2}} \sum_{\alpha=1}^{3} h_{i \alpha} M_{\psi_\alpha}y^{*}_{j\alpha}\,, \\
& \sim 0.04~\text{eV} \left( \frac{\lambda_8}{10^{-4}}\right) \left( \frac{v_{S}}{200\,  \text{GeV}}\right) \left( \frac{M_{\psi_\alpha}}{50\,\text{GeV}}\right) \left( \frac{2\, \text{TeV}}{m_{\chi}}\right)^{2} \left( \frac{h_{i \alpha} y_{j\alpha}}{10^{-4}}\right)\,. 
\end{align}
In this way, in addition to the loop suppression it is possible to 
have further suppression in the neutrino mass matrix for small values
of either $v_S$, $\lambda_8$ or $h_{i \alpha} y_{j\alpha}$.

\begin{figure}[t]
\centering
\includegraphics[scale=0.6]{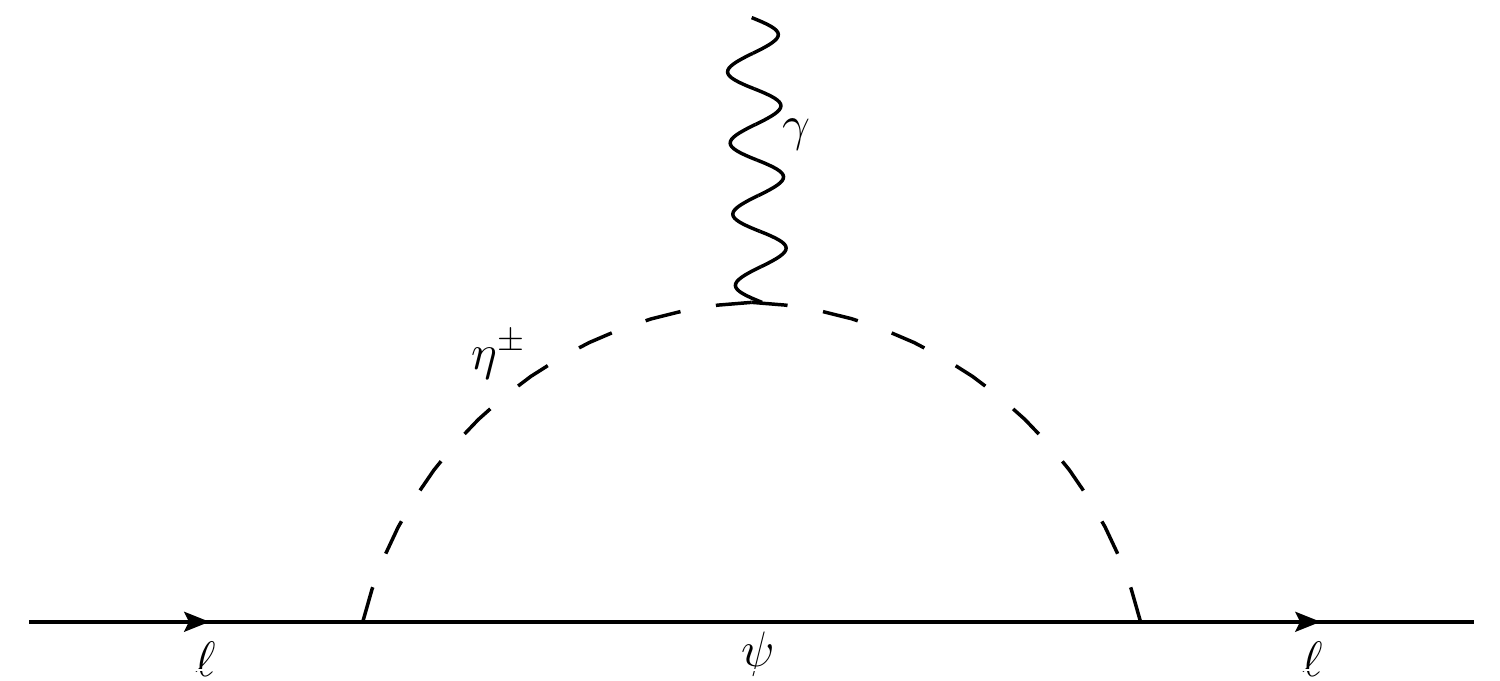}
\caption{Feynman diagram for the processes $\ell_{i} \to \ell_{j} \gamma$}
\label{fig:LFV}
\end{figure}

On the other hand, the $h_{i\alpha}$ Yukawa interaction in Eq.~\eqref{eq:LagY} leads to charged lepton flavor violation (CLFV) processes induced at one-loop level and mediated by the charged scalars $\eta^{\pm}$ as the ones shown in Fig.~\ref{fig:LFV} for the $\ell_i\to\ell_j\gamma$ type. 
By using the current experimental constraint on $\operatorname{Br}(\mu \to e\gamma)<5.7 \times 10^{-13} $~\cite{Adam:2013mnn} and for the case $m_\chi^2=m_{\eta^{\pm}}^{2} \gg M_{\psi_{a}}^{2}$
 we can obtain an upper bound for the product of Yukawa couplings
\begin{align}
    \left| \sum_{\alpha}h_{2 \alpha} h_{1\alpha}^{*} \right| \lesssim 0.02 \left(\frac{m_\chi}{2\,\text{TeV}} \right)^{2}.
\end{align}
It is worth noticing that $y_{j\alpha}$ is not constrained by the non observation of CLFV processes. 

\section{Dark matter}
\label{sec:DM}
From the model charge assignment in Table~\ref{tab:partcont} we have
that a residual $Z_2$ symmetry is left over after the
$\operatorname{U}(1)_{\mathcal{D}}$ symmetry breaking, with the
particles circulating the one-loop neutrino mass diagram (see
Fig.~\ref{fig:zee}) and $\nu_{R3}$ being odd under it whereas
$\nu_{R1}, \nu_{R2}$, $S$ and all the SM particles being even. Thus the lightest electrically-neutral $Z_2$-odd
particle becomes a dark matter candidate. In other words, this model
also provides a solution to the DM puzzle via either fermion
($\psi_1$) or scalar ($\chi_{1}$) dark 
matter\footnote{Note that $Z'$ cannot constitute a DM candidate due
  to the instability associated to the $Z'\to\bar{\nu}_{R3}\nu_{R3}$
  decay channel, which cannot be kinematically closed, and that neither $\psi_1$ and $\chi_{1}$ can decay into $\nu_{R3}$ since they do not share Yukawa interactions (see Eq.~(\ref{eq:LagY})).}.

Since $\psi_{1}$ is a singlet under the SM gauge group its thermal relic density is controlled by the Yukawa couplings and $\operatorname{U}(1)_{\mathcal{D}}$ gauge interactions. 
The scenario where $\psi_{1}$ self-annihilates dominantly through $h_{i\alpha}$-mediated interactions resembles the very well known scotogenic model \cite{Ma:2006km},  
where sizable $h_{i\alpha}$ are required to reproduce the correct DM abundance, which in turn leads to a mild tension with experimental upper bounds on the rates for rare charged lepton decays~\cite{Kubo:2006yx,Sierra:2008wj,Ibarra:2016dlb}. 
On the other hand, when the $Z'$ portal \cite{Langacker:2008yv,Arcadi:2013qia} is the main gate to visible sector, $\psi_1$ largely annihilates into neutrinos in such a way the observed DM abundance can be reproduced without entering in conflict with the DM searches, which follows from the fact that the $Z'$ does not couple to quarks and charged leptons
(see~\cite{Arcadi:2013qia,Alves:2013tqa,Alves:2015pea,Alves:2016cqf,Blanco:2019hah} for phenomenological studies on $Z'$-mediated Majorana DM). 

The elastic scattering of $\psi_1$ particles off nuclei may occur via different mediators: inert scalars ($\sigma,\eta$), SM Higgs (through the mixing with $S$) and $Z'$. When the inert scalars mediate  DM nucleon scatterings the corresponding spin independent cross section is loop suppressed due to their quark-phobic nature \cite{Ibarra:2016dlb}, whereas the scatterings via the $Z'$ require necessarily a kinetic mixing (since no SM fermion couples directly to $Z'$) and lead to a spin-dependent cross section. Lastly, the scatterings via the Higgs exchange are also suppressed by the mixing parameter $\lambda_1$, which is required to be small to correctly reproduce the SM Higgs phenomenology. 
All in all, it is feasible to expect that the rates for the elastic scattering of $\psi$ particles off nuclei lie well below the sensitivity of present direct detection experiments. 

In contrast to fermion DM, the scalar DM candidate has additional interaction terms to the Yukawa and gauge interactions. This entails that the later ones can be used to alter the relic density predictions in scenarios with mixed scalar DM. Since in the present model the CP-even and CP-odd neutral $Z_2$-odd particles are mass degenerate, we have the scenario of singlet-doublet complex DM~\cite{Kadastik:2009dj,Belanger:2012vp} where the DM candidate is a mixture of a complex singlet \cite{McDonald:1993ex} and a $SU(2)_L$ doublet \cite{Deshpande:1977rw,Barbieri:2006dq}. It follows that for negligible Yukawa and gauge interactions there are two DM mass regions that allow us to properly reproduce the relic abundance, one corresponds the Higgs funnel region and the second one demands masses above $100$ GeV~\cite{Kakizaki:2016dza}. 
Regarding direct detection signals, we expect similar scattering rates as those in the complex scalar singlet model \cite{Cline:2013gha,Wu:2016mbe}. 

\section{Beyond the minimal model: cosmological and collider constraints}
\label{sec:CosmoConstraints}
It turns that the $Z'$ portal also allows to probe the model through modifications on the cosmological history of the Universe, namely, via additional contributions to the effective number of relativistic degrees of freedom $N_{\text{eff}}$~\cite{Dolgov:2002wy}. In this model, these contributions arise from the presence of the right-handed neutrinos and may be expected to be sizeable precisely due to the large $\operatorname{U}(1)_{\mathcal{D}}$~charges of the $\nu_R$'s.  Nevertheless, since the right-handed neutrinos do not couple directly to the rest of the SM particles (see Eq.~(\ref{eq:LagY}))  they  decouple early enough (when the DM does or even before) from the thermal bath and, therefore, modify the SM prediction for $N_{\text{eff}}$ with an extra contribution at most of $\sim0.2$~\cite{Chacko:2015noa}.

With the aim of extending the thermalization period of the right-handed neutrinos with the
primordial plasma in the early Universe, we consider an modification of
the previous setup by using a general anomaly free Abelian gauge
symmetry with generation-independent charge assignments for the
SM fermions, $\operatorname{U}(1)_{X}$.
In the Appendix we show that the solutions to the anomaly cancellation conditions
allow us to write the $X$-charges of the SM fields in terms
of two parameters~\cite{Appelquist:2002mw,Campos:2017dgc,Allanach:2018vjg,Das:2019pua}, that we choose as
\begin{align}
  X(r,l)=r R - l Y\,,
\end{align}
where $R$ is the generator of $\operatorname{U}(1)_R$ (the gauge Abelian symmetry where only the right-handed SM fermions have non-vanishing $X$-charges), $Y$ is the hypercharge, $l$ is the $X$-charge of the lepton doublets, and $r$ parametrizes the contribution to the linear and mixed gauge-gravitational anomalies of any extra set of chiral fermions, as given in Eq.~\eqref{eq:anolam} of the Appendix.
In this way, if only extra vector-like fermions are allowed beyond the SM ($r=0$) the solution must be proportional to hypercharge.
Without lost of generality, we can write the solutions in terms of just one parameter~\cite{Jenkins:1987ue,Oda:2015gna,Okada:2018tgy}, that we choose to be $l$  after fix $r=1$. Then, as shown in the Appendix, the full family of solutions for a fixed $l$ can be obtained after  rescaling all the $X$-charges by a factor $r$. In particular, the fermiophobic~\cite{Campos:2017dgc,Bertuzzo:2018itn,Bertuzzo:2018ftf} solution used in the previous sections, $\operatorname{U}(1)_{\mathcal{D}}$, corresponds to the rescaling $r=0$ of the $\operatorname{U}(1)_R$ solution: $\mathcal{D}=X(0,0)$. 

The one parameter solution is shown in  column $\operatorname{U}(1)_X$
of Table~\ref{tab:partcont3}.
In order to analyse the phenomenology, we fix $l$ to recover some already studied
Abelian gauge groups $X=B-L,R,D,G,$ as defined in Ref.~\cite{Campos:2017dgc}\footnote{We change $\operatorname{U}(1)_B$ for the more suitable name of $\operatorname{U}(1)_R$~\cite{Jana:2019mez}.}. The last column corresponds to the rescaling with $r=0$ of $\operatorname{U}(1)_R$.

Since we are interested here in keeping the exotic set of $X$-charges
charges $\{\mp 4$,~$\mp 4$,~$\pm 5\}$ in such a way that when assigned
to the right-handed neutrinos the tree level Dirac and Majorana masses
can be forbidden~\cite{Calle:2018ovc}, the set of Majorana mediators
$\psi_{R\alpha}$ which realize the dimension-6 operator at one-loop
 would spoil the anomaly cancellation
condition.
In view of that $\psi_{R\alpha}$ have necessarily nonzero $X$-charges (see the Appendix), we further add an extra set of chiral fermions in 
such a way the full set heavy fermions  do not affect
the anomaly cancellation (their charges cancel each other in a vector like way). 
The resulting charge assignment is displayed in
Table~\ref{tab:partcont3}, where 
the fields $\xi_{L \alpha}$ constitute the new set of chiral fermion that guarantee the anomaly cancellation.
In this way we end up with a model with four Majorana fermions, two more than in the minimal solution, since third generation of chiral fermions is not needed to cancel the $\operatorname{U(1)}_X$ anomalies (alternatively we may consider another simple setup by adding a single set of $\psi_R$ and $\xi_L$ and two set of scalars $\eta_\alpha$, $\sigma_{\alpha}$~\cite{Reig:2018mdk})\footnote{
Seeing that the two additional left-handed fields $\xi_{L\alpha}$,  with a $X$-charge $r=3/4$,  have both Dirac and Majorana mass terms $\overline{\xi_{L \alpha}}\psi_{R \beta}$ and $ \overline{\xi_{L\alpha}^c }\xi_{L \beta} \langle S\rangle$, the full set of Majorana fields are massive and heavy. 
Moreover, because the $\operatorname{U}(1)_{X}$ left out a remnant $Z_2$ discrete symmetry which guarantees the stability of the lightest $Z_2$-odd particle,  we have that the lightest of these Majorana fields may play the role of DM candidate.}.

\begin{table}
  \centering
  \begin{tabular}{|c|c|c|c|c|c|c|c||c|}
    \hline  
    Fields     & $\operatorname{SU}(2)_L$ &  $\operatorname{U}(1)_Y $ & $\operatorname{U}(1)_{X}$& $\operatorname{U}(1)_{B-L}$& $\operatorname{U}(1)_R$& $\operatorname{U}(1)_D$& $\operatorname{U}(1)_G$ & $\operatorname{U}(1)_{\mathcal{D}}$\\ \hline
$L $     & $\boldsymbol{2}$ & $-1$ & $l$      &  $-1$&    $0$ &  $-3/2$&  $-1/2$ & $0$ \\    
$d_R $   & $\boldsymbol{1}$ & $-2/3$ & $1+2l/3$ &  $1/3$&    $1$&  $0$&    $2/3$ & $0$\\
$u_R $   & $\boldsymbol{1}$ & $+4/3$ & $-1-4l/3$&  $1/3$&   $-1$&  $1$&   $-1/3$ & $0$ \\
$Q $     & $\boldsymbol{2}$ & $1/3$ & $-l/3$   & $1/3$&    $0$&  $1/2$&  $1/6$ & $0$ \\
$e_R $   & $\boldsymbol{1}$ & $-2$   & $1+2l$   &  $-1$&    $1$ &  $-2$&  $0$ & $0$\\\hline
$H $     & $\boldsymbol{2}$ & $1$  & $-1-l$   &  $0$&    $-1$ &  $1/2$&  $-1/2$ & $0$ \\
$\eta$   & $\boldsymbol{2}$ & $1$ & $3/4-l$  & $7/4$& $3/4$ &$9/4$&$5/4$ & $1$ \\
$S$      & $\boldsymbol{1}$ & $0$    & $3/2$    & $3/2$&  $3/2$ & $3/2$& $3/2$ & $2$\\
$\sigma$ & $\boldsymbol{1}$ & $0$    & $13/4$   & $13/4$&  $13/4$ & $13/4$& $13/4$ & $3$\\\hline
$\nu_{Ri}$& $\boldsymbol{1}$ & $0$   & $-4$& $-4$&  $-4$ & $-4$& $-4$ & $-4$\\
$\nu_{R3}$& $\boldsymbol{1}$ & $0$   & $+5$& $+5$&  $5$ & $5$& $5$ & $5$\\
$\psi_{R\alpha}$  & $\boldsymbol{1}$ & $0$& $3/4$ & $3/4$&  $3/4$ & $3/4$& $3/4$ & $1$ \\\hline
$\xi_{L\alpha}$   & $\boldsymbol{1}$ & $0$ & $3/4$& $3/4$ &  $3/4$ & $3/4$& $3/4$ &$-$\\\hline
  \end{tabular}
  \caption{General one-parameter solution with some examples of rational solutions
    ($X=B-L,R,D,G$ and $\mathcal{D}$) for the radiative type-I seesaw realization
    of the effective operator $\mathcal{O}_{6D}$ for Dirac neutrino masses. The
    last column  corresponds to the solution in Table~\ref{tab:partcont}.}
    \label{tab:partcont3}
\end{table}

Accordingly, we may expect that 
the three right-handed neutrinos within the not-so minimal model
contribute to the radiation energy density of the Universe, since now   the interaction between the $Z'$ with the SM fermions  opens up the possibility to thermalize them with the primordial plasma.   
In other words, this leads to a modification in the relativistic degrees of freedom as~\cite{Anchordoqui:2012qu,Anchordoqui:2011nh}
\begin{align}
    \Delta N_{\text{eff}} = N_{\text{eff}} - N^{\text{SM}}_{\text{eff}} = N_{\nu_R} \left( \frac{T_{\nu_{R}}}{T_{\nu_{L}}} \right)^{4} = N_{\nu_R} \left( \frac{g(T^{\nu_{L}}_{\text{dec}})}{g(T^{\nu_{R}}_{\text{dec}})} \right)^{4/3},
\end{align}
where $N_{\nu_R}$ is the number of right-handed neutrinos with the same $X$ charge and $g(T)$ is the number of relativistic degrees of freedom at a temperature $T$ in the SM~\cite{Borsanyi:2016ksw}. The decoupling temperature of the SM neutrinos is $ T^{\nu_{L}}_{\text{dec}} \approx 2.3 $ MeV, when $g(T^{\nu_{L}}_{ \text{dec}}) = 43/4$ corresponding to the three $\nu_{L}$, $e^{\pm} $ and the photon~\cite{Kolb:1990vq,Enqvist:1991gx}. Since the interaction of the right-handed neutrinos with the SM is only mediated by the gauge boson $Z^{\prime} $, the corresponding rate can be cast as~\cite{SolagurenBeascoa:2012cz}
\begin{align}
    \Gamma_{\nu_R} (T) &= n_{\nu_R}(T) \langle \sigma(\overline{\nu_{R}} \nu_{R} \to \overline{f} f) \upsilon \rangle \, \nonumber\\
    &= \frac{g^{2}_{\nu_R}}{n_{\nu_R}(T)} \int \frac{d^{3} p}{(2 \pi)^{3}} f_{\nu_R}(p) \int \frac{d^{3} k}{(2 \pi)^{3}} f_{\nu_R}(k) \sigma_{f}(s) \upsilon\,,
\end{align}
where $f_{\nu_R}(k)=1/(e^{k/T}+1)$ is the Fermi-Dirac distribution, $g_{\nu_R} = 2$, $\upsilon = 1-\cos{\varphi}$ is the Moller velocity, $s = 2 p k (1-\cos{\varphi})$, $p$ and $k$ are the momenta of the particle with $\varphi$ the angle between them, and the number density of right-handed neutrinos is given by
\begin{align*}
n_{\nu_R}(T) = g_{\nu_R} \int \frac{d^{3} k}{(2 \pi)^{3}} f_{\nu_R}(k)\,.
\end{align*}
The cross section for the case of a heavy mediator ($T^{\nu_R}_{\text{dec}} \ll M_{Z^{\prime}}$), with $s \ll M_{Z^{\prime}}$ and neglecting the fermion masses in the final state reads~\cite{Barger:2003zh}
\begin{align}
    \sigma_{f}(s) \approx N^{C}_{f} \frac{s}{12 \pi} \left( \frac{g^{\prime}}{M_{Z^{\prime}}} \right)^{4} q^{2}_{\nu_R} (q^{2}_{f_L} + q^{2}_{f_R})\,,
\end{align}
where $N^{C}_{f}=1(3)$ for leptons (quarks),
and $q_{f}$ is the $X$-charge of the SM fermion.  
Accordingly, the interaction rate takes the form
\begin{align}
    \Gamma_{\nu_{R}}(T) = \frac{49 \pi^{5} T^{5}}{97200 \zeta(3)} \left( \frac{g^{\prime}}{M_{Z^{\prime}}} \right)^{4} \sum_{f} N^{C}_{f} q^{2}_{f}\,.
\end{align}
In this expression the sum is performed over all SM fermions that are in thermal equilibrium with the plasma at temperature $T$. To estimate the contribution to the relativistic degrees of freedom from right-handed neutrinos, it is necessary to calculate the decoupling temperature of the right-handed neutrinos ($T^{\nu_R}_{\text{dec}}$). The latter occurs when the interaction rate $\Gamma_{\nu_{R}}(T)$ drops below the rate of expansion of the Universe, $\Gamma(T^{\nu_R}_{\text{dec}}) = H(T^{\nu_R}_{\text{dec}})\,, $
with $H(T)= [4 \pi^{3} G_{N}(g(T) + 21/4)/45]^{1/2} T^{2}\,.$
Note that the factor $21/4$ corresponds to the contribution of the right-handed neutrinos to the relativistic degrees of freedom. 

\begin{figure}[t]
\centering
\includegraphics[scale=1]{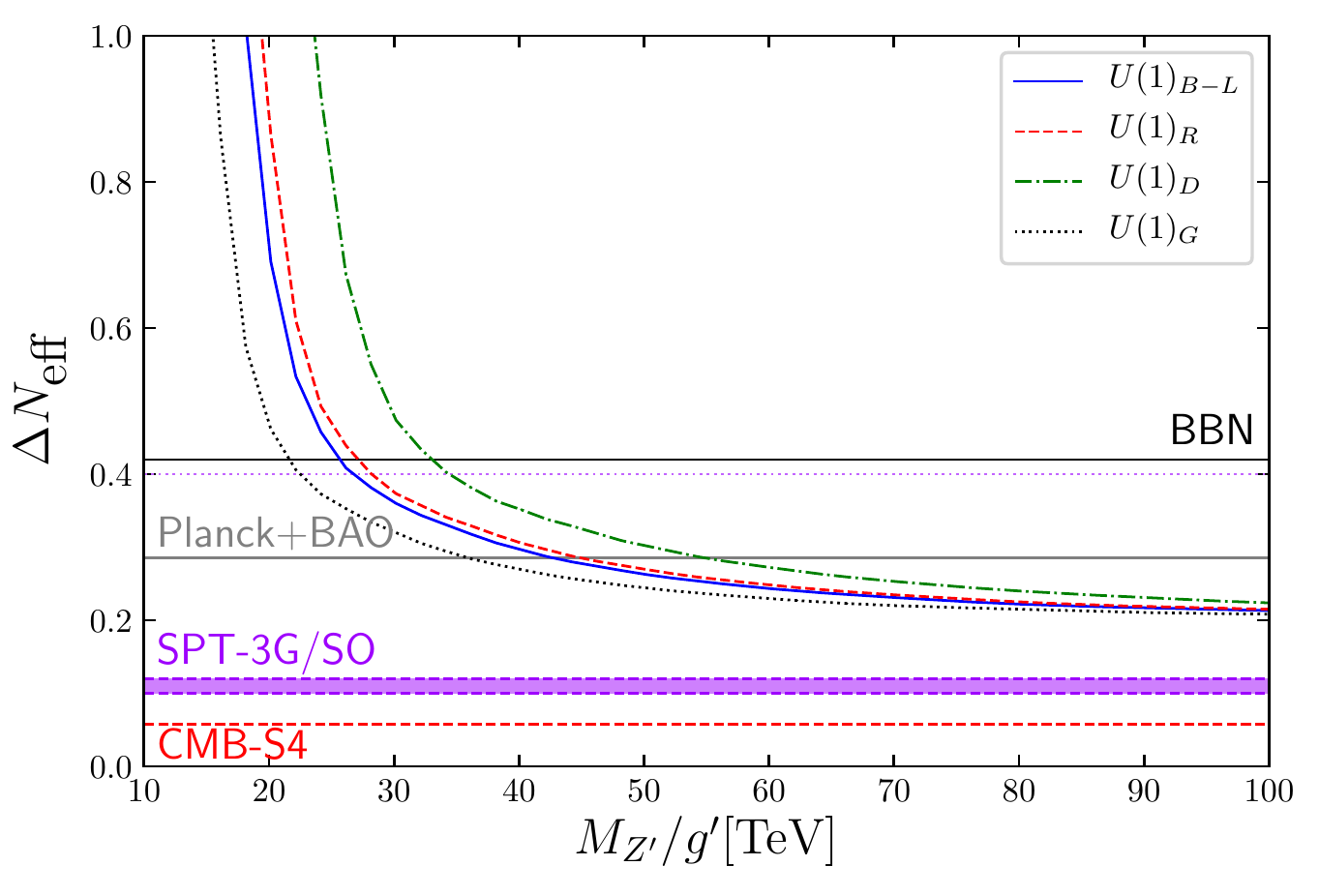}
\caption{Contribution to the number of extra relativistic degrees of freedom ($\Delta N_{\text{eff}}$) in function of $M_{Z^{\prime}}/g^{\prime}$. The region above the solid gray line is excluded by the measurements at $2\sigma$ reported by the PLANCK Collaboration~\cite{Aghanim:2018eyx}. For comparison purposes  the upper bound at $2\sigma$ (black line) obtained from the BBN analysis~\cite{Pitrou:2018cgg} is also shown.  The horizontal dashed lines show the projected sensitivity of the future experiments SPT-3G/SO~\cite{Benson:2014qhw,Ade:2018sbj} and CMB-S4~\cite{Abitbol:2019nhf}. $\Delta N_{\text{eff}}=0.4$ represents the extra contribution  required to relieve the tension on the inferred $H_0$ values from  high- and low-redshift observations \cite{Bernal:2016gxb,Mortsell:2018mfj}. }
\label{fig:Neff}
\end{figure}

The results for $\Delta N_{\text{eff}}$ as a function of $M_{Z^{\prime}}/g^{\prime}$ are displayed in Fig.~\ref{fig:Neff}, where it can be observed that for small ratios $M_{Z'}/g'$  the variation in the number of relativistic degrees of freedom of the new species is large ($\Delta N_{\text{eff}}\gtrsim0.3$ corresponds to decoupling temperatures $T^{\nu_R}_{\text{dec}}\lesssim2$ GeV) and vice versa.
It follows that for the $\operatorname{U}(1)_{G}$ model the Planck+BAO upper bound at $2\sigma$ (solid gray line) demands that $M_{Z^{\prime}}/g^{\prime} \gtrsim 36$ TeV while for the $\operatorname{U}(1)_{D}$ model a more stringent lower bound applies $M_{Z^{\prime}}/g^{\prime} \gtrsim 55$~TeV (these bounds become slightly weaker once the Planck+BAO+$H_0$ combination~\cite{Aghanim:2018eyx} is considered, but in such a case the BBN bound from the primordial abundances of light elements would rule~\cite{Pitrou:2018cgg}). 
In other words, bearing in mind that $M_{Z'}/g'\approx 3v_S/2$\footnote{Because of the LEP constraint (see below) the mixing between the $Z'$ and the SM $Z$ boson is negligible~\cite{Carena:2004xs}.} we have that the energy scale of the $\operatorname{U}(1)_{X}$ symmetry breaking must be at least $\sim24$ TeV 
(note that the  $\operatorname{U}(1)_X$ having the lowest  $M_{Z'}/g'$ ratio features a X charge $l=-6/11\approx-1/2$). 
Similar constraints would apply to all gauged and anomaly free  $\operatorname{U}(1)_X$ extensions of the SM with Dirac neutrino masses, where light right handed neutrinos are associated  to the solution $(-4,-4,+5)$~\cite{Ma:2014qra,Ma:2015mjd,Ma:2015raa,Bonilla:2018ynb,Calle:2018ovc,Bonilla:2019hfb}.
On the other hand, it is remarkable the fact the next generation of CMB experiments~\cite{Benson:2014qhw,Ade:2018sbj,Abitbol:2019nhf} has the potential to entirely probe all the not-so minimal models~\cite{Baumann:2017gkg,Abazajian:2019oqj}.

\begin{figure}[t]
    \centering
    \includegraphics[scale=1.0]{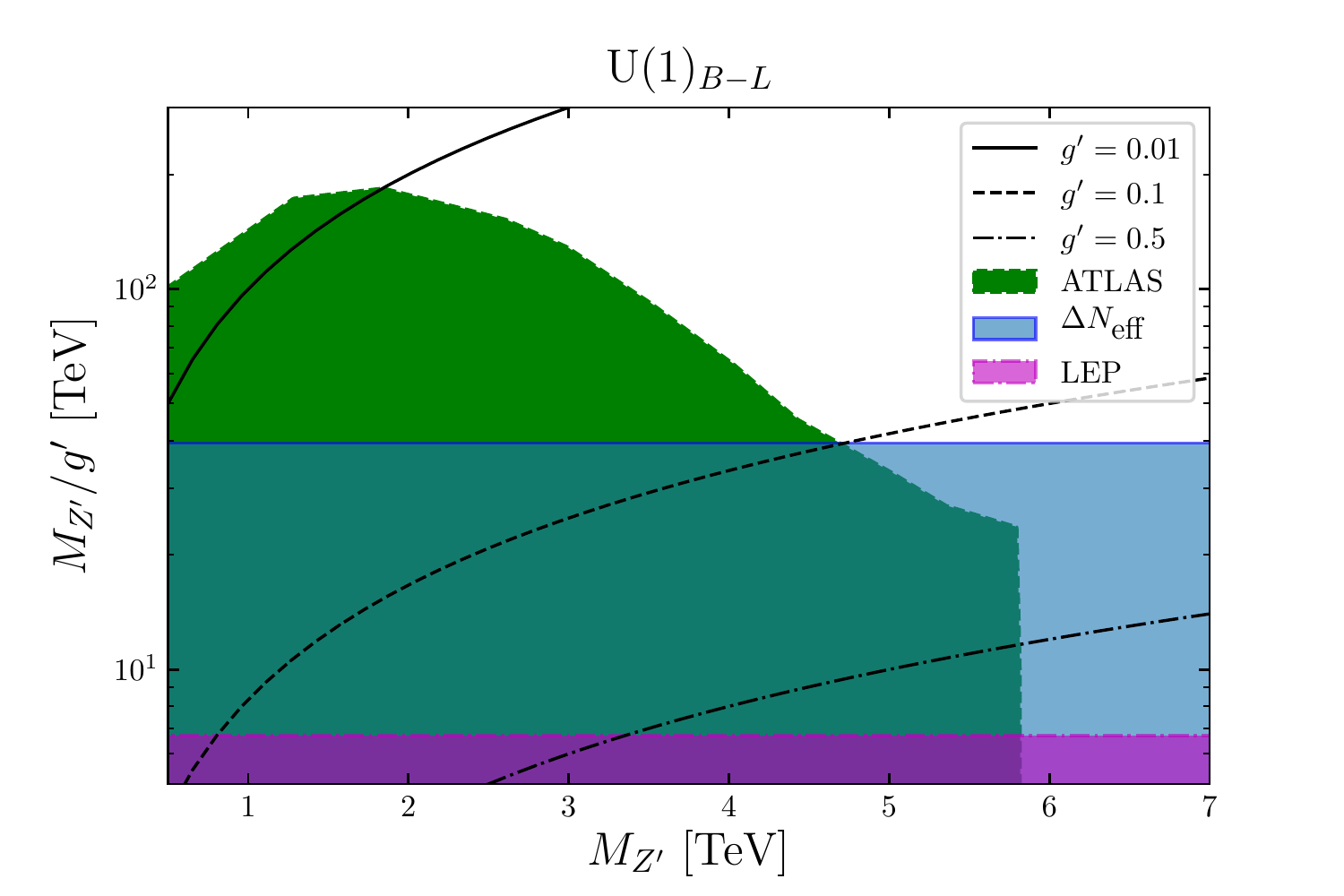}
    \caption{Collider and cosmological constraints for the $\operatorname{U}(1)_{B-L}$ model, and contours of constant $g'$ in the plane of $q_S v_S$ as a function of $M_{Z'}$ with $q_S=3/2$.}
    \label{fig:u1blc}
\end{figure}

Regarding collider searches, the recasting of the latest ATLAS results for the search of dilepton resonances using $139\ \text{fb}^{-1}$~\cite{Aad:2019fac} was done
in Ref.~\cite{Chiang:2019ajm} for the $\operatorname{U}(1)_{B-L}$ model.
The green (upper) region in Fig.~\ref{fig:u1blc} shows the excluded region at $95\%\ \text{C.L.}$ To ease the comparison with other results, we show the exclusion as function of $M_{Z'}/g^{\prime}$. In particular, the limit from LEP for $\operatorname{U}(1)_{B-L}$ model is~\cite{Carena:2004xs,Heeck:2014zfa,Das:2015nwk}\footnote{The constraint as function of the $X$-charge $h=-1-l$ is given in Ref.~\cite{Okada:2016tci}. }
\begin{align} 
  M_{Z'}/g^{\prime}>6.7\ \text{TeV}\,,&&\text{at $95\%$ C.L }\,,
\end{align}
which is obtained from the search for effective four-lepton operators and is valid for $M_{Z'}\gg 200\ \text{GeV}$. This constraint corresponds to the excluded magenta (lower) region of Fig.~\ref{fig:u1blc} and start to be relevant for $M_{Z'}>5.8\ \text{TeV}$.

The constraint of $\Delta N_{\text{eff}}$ for $\operatorname{U}(1)_{B-L}$ is shown in the blue (middle) region and start to be better than current ATLAS limit for $M_{Z'}\gtrsim 4.8\ \text{TeV}$. In the figure we also show contours of constant $g'$ at $0.01$, $0.1$ and $0.5$ with the solid, dashed and dot-dashed lines respectively. We can see that $\Delta N_{\text{eff}}$ start to constraint $g'$ for values larger than $0.1$ in $\operatorname{U}(1)_{B-L}$.    Similar restrictions can be obtained for the other models quoted in Fig.~\ref{fig:Neff}.

\section{Conclusions}
The mechanism behind the neutrino mass generation and the Dirac/Majorana character of massive neutrinos still remain a conundrum despite the clear understanding of the neutrino oscillation pattern and the great experimental efforts behind the neutrinoless double-beta decay. Alongside this is the fact that there are no clues on the nature of the DM particle and the properties of the dark sector it belongs. 

In view of this we have proposed a new mechanism for Dirac neutrino masses which making use of heavy Majorana mediators generate nonzero masses at one-loop level. 
The model presented in this paper is a one-loop realization of the dimension 6 operator $\overline{L} \tilde{H} \nu_R S^2\,$ and enters into the list of Dirac radiative type-I seesaw models with the novel feature that it involves Majorana mediators rather than Dirac mediators. 
The Diracness of the massive neutrinos is protected by only one extra $\operatorname{U}(1)_X$ gauge symmetry, which in turn ensures the stability of the lightest particle mediating the one-loop neutrino mass diagram. 
In this way the model offers in a non trivial way a the solution to both neutrino and DM puzzles.
Moreover, we have shown that going beyond the minimal model other interesting phenomenological aspects arise such that possible signals at colliders and new contributions to the number of extra relativistic species. In particular, the  current upper bound on $\Delta N_{\text{eff}}$ reported by PLANCK points to large ratios $M_{Z'}/g'\gtrsim 40\ \text{TeV}$. Future cosmic microwave background experiments may probe all the no minimal models presented here.

\section*{Acknowledgments}
Work supported by Sostenibilidad-UdeA and the UdeA/CODI Grant
2017-16286, and by COLCIENCIAS through the Grant 111577657253. D.R
would also like to thank the IIP-UFRN, for its support,
hospitality and the nice and stimulating atmosphere during the
Visiting Professors program where part of this work was completed. 
O.Z. acknowledges the ICTP Simons associates program.  
\appendix
\section{$\operatorname{U}(1)_X$ anomaly cancellation}

We use $f$ ($f$) to denote the general $\operatorname{U}(1)_X$ generation-independent charge assignments of the field $f_R$ ($F_L$). The three linear anomalies in $\operatorname{U(1)_X}$~\cite{Campos:2017dgc}
\begin{align}
  \label{eq:anolin}
  \left[\mathrm{SU}(3)_{C}\right]^{2} \mathrm{U}(1)_{X} :& &
[3 u+3 d]-[3 \cdot 2 q]=&0, \nonumber\\
\left[\mathrm{SU}(2)_{L}\right]^{2} \mathrm{U}(1)_{X} :&&
-[2 l+3 \cdot 2 q]=&0, \nonumber\\
\left[\mathrm{U}(1)_{Y}\right]^{2} \mathrm{U}(1)_{X} :&&
{ \left[(-2)^{2} e+3\left(\tfrac{4}{3}\right)^{2} u+3\left(-\tfrac{2}{3}\right)^{2} d\right]-\left[2(-1)^{2} l+3 \cdot 2\left(\tfrac{1}{3}\right)^{2} q\right]}=&
 0,
  \end{align}
allows to express three $X$-charges in terms of the other two
\begin{align}
  u=&-e+\frac{2l}{3}\,,& d=& e-\frac{4l}{3}\,,& q=& -\frac{l}{3}\,.
\end{align}
The quadratic anomaly condition is automatically satisfied, while the mixed gauge-gravitational and cubic anomalies depend of any extra singlet quiral fermions of zero hypercharge, like the right-handed counterpart of the Dirac neutrinos. For $N$ extra quiral fields with $X$-charge $n_{\alpha}$, these conditions read
\begin{align}
  \left[\text{Grav}\right]^{2} \mathrm{U}(1)_{X}:\ \sum_{\alpha=1}^N n_{\alpha}+3 (e-2l)=&0\,, &  \left[\mathrm{U}(1)_{X}\right]^{3}:\ \sum_{\alpha=1}^N n_{\alpha}^3+3 (e-2l)^3=&0\,. &
\end{align}
We choose the solutions with $r\equiv e-2l$, such that
\begin{align}
  \label{eq:anolam}
  \sum_{\alpha=1}^{N} n_{\alpha}=&-3 r\,,   & \sum_{\alpha=1}^{N} n_{\alpha}^3=&-3 r^3\,.
\end{align}
The full set of anomaly free SM $X$-charges in terms of two parameters~\cite{Appelquist:2002mw,Campos:2017dgc,Allanach:2018vjg} that we choose as $l$ and $r$, is just
\begin{align}
  u=&-r-\frac{4l}{3}\,,&d=&r+\frac{2l}{3}\,,&q=&-\frac{l}{3}\,,&e=&r+2l\,,&h=&-r-l\,.
  \label{Eq:SMCharges}
\end{align}
where the condition in the charged lepton Yukawa couplings have been used to fix $h$, and is automatically consistent with the conditions in the quark Yukawa couplings. By setting $l=0$ in the previous equations, we can define the Abelian symmetry in which only the right-handed charged fermions have non-vanishing $X$-charges as $\operatorname{U}(1)_R$. Then the general anomaly free two-parameter solution can be written as
\begin{align}
  X(r,l)=r R- l Y\,.
\end{align}

If we now change $f\to f'= f/r$ for all the charged fermion $X$-charges~\cite{Allanach:2018vjg}, the
first set of anomaly cancellation conditions Eq.~\eqref{eq:anolin} remains
invariant, and without lost of generality it is always possible to
normalize the solutions such that the last set Eq.~\eqref{eq:anolam} is
just
\begin{align}
  \label{eq:anolamnor}
   \sum_{\alpha=1}^{N} n_{\alpha}^{\prime}=&-3\,,   & \sum_{\alpha=1}^{N} n_{\alpha}^{\prime\, 3}=&-3\,.
\end{align}
For example, the solution with $r=3$: $n_{\alpha}=\left( -2,-2,-4,-1 \right)$~\cite{Appelquist:2002mw} can be easily normalized to the form in Eq.~\eqref{eq:anolamnor} with $f\to f/3$
to $n_{\alpha}'=\left( -2/3,-2/3,-4/3,-1/3 \right)$
as used in Ref.~\cite{Patra:2016ofq}. In this way, without lost of generality, we will work with the normalized solution in terms of a single parameter~\cite{Jenkins:1987ue,Oda:2015gna,Okada:2018tgy} that we choose to be $l$, by setting $r=1$ as summarized in column $\operatorname{U}(1)_X$ of Table~\ref{tab:partcont3}, which is just
\begin{align}
  X(l)=R-l\, Y\,.
\end{align}
In particular, this includes the solution  $n_{\alpha}=(-4,-4,+5)$~\cite{Appelquist:2002mw}. 
In general, we have that for $\nu=n_1=n_2$, the extra fermion inside the one-loop neutrino mass diagram in Fig.~\ref{fig:zee}, must have charges
\begin{align}
  \psi=&-\frac{\nu+1}{4}\,,&\eta=&-\frac{\nu+1}{4}-l\,,&\sigma=&\frac{1-3\nu}{4}\,.
\end{align}
The case for $\nu=-4$ is also displayed in Table~\ref{tab:partcont3}.

\bibliographystyle{apsrev4-1long}
\bibliography{susy}

\end{document}